\documentclass[aps,prc,showpacs,twocolumn]{revtex4}
\usepackage{graphicx}
\usepackage{dcolumn}
\usepackage{bm}
\usepackage{longtable}
\usepackage{hyperref}
\usepackage{natbib}
\usepackage{amsmath}
\begin{document}
\title{Partial restoration of isospin symmetry for neutrinoless double beta decay in the deformed nuclear system of $^{150}$Nd }
\author{Dong-Liang Fang$^{a}$, Amand Faessler$^{b}$ and Fedor Simkovic$^{c,d}$}
\affiliation{$^a$College of Physics, Jilin University, Changchun, Jilin 130012, China}
\affiliation{$^b$Institute of Theoretical Physics, University of Tuebingen,  D-72076 Tuebingen, Germany}
\affiliation{$^c$INR, 141980 Dubna, Moscow Region, Russia}
\affiliation{$^d$Comenius University, Physics Department, SK-842 15 Bratislava, Slovakia}
\begin{abstract}
In this work, we calculate the matrix elements for the  $0\nu\beta\beta$-decay  of $^{150}$Nd using the deformed pn-QRPA method. We adopted the approach introduced by Rodin and Faessler [Phys. Rev. C84, 014322 (2011)] and by Simkovic {\it et. al.} [Phys. Rev. C87,045501(2013)] to restore the isospin symmetry by enforcing $M^{2\nu}_F=0$. We found that with this restoration, the Fermi matrix elements are reduced in the strongly deformed $^{150}Nd$ by about 15 to 20\%, while the more important Gamow-Teller matrix elements remain the same. The results of an enlarged model space are also presented. This enlargement increases the total (Fermi plus Gamow-Teller) matrix elements by less than 10\%.
\end{abstract}
\maketitle
The neutrinoless double beta decay (hereafter $0\nu\beta\beta$-decay), if it exists, is one of the rarest processes in our universe. Its discovery could lead to the dawn of a era for new physics beyond the Standard Model. Various experiments search or are proposed to look for  this exotic process. Currently, about a dozen isotopes have been confirmed to decay by the two-neutrino double beta decay ($2\nu\beta\beta$) mode (for a review see \cite{Bar10}). They are also good candidates for $0\nu\beta\beta$-decay. Of these isotopes $^{150}$Nd has a relative small $2\nu\beta\beta$ half-life as shown in Table I of ref. \cite{Bar10}. This decay has also a large Q value (Q = 3.367 MeV, see table I of ref. \cite{Bilenky}) . Recent calculations in ref. \cite{KI12} show, that it also has the largest phase space for the light Majorana neutrino mechanism -- the most probable mechanism for this decay. Thus this decay system is a promising candidate with possibly the shortest $0\nu\beta\beta$-decay half-life. To confirm this assumption, one needs to investigate further this process. Compared to other $\beta\beta$-active isotopes, $^{150}$Nd is supposed to be strongly deformed. This poses difficulties for exact shell model calculations. So for this nucleus one needs to resort to other methods, {\it e. g.:} the Projected HFB \cite{RCCRH10}, the Interacting Boson Model (IBM) \cite{BKI15} and Energy Density Functional (EDF) (non-relativistic ref. \cite{RM10} and relativistic ref.  \cite{YSHRM15}). These methods calculate the ground states of the initial and final nuclei of the double beta decay, and then use closure for the intermediate states with an averaged energy denominator to obtain the nuclear matrix element. In this way one needs not to calculate the wave functions of the intermediate nucleus ( $^{150}$Pm). Another category of methods used extensively is the Quasi-particle Random Phase Approximation (QRPA), which calculates explicitly the intermediate states. According to the mean fields and interactions, different versions of QRPA are in use. Our group uses QRPA with realistic forces \cite{FFRS11}. 
There exist also QRPA calculations with the Skyrme interaction \cite{ME13} and its variations \cite{Ter15}. 

Recently, a new formalism \cite{Rodin, SRFV13} with restored isospin symmetry has been developed by the T\"ubingen group for QRPA calculations. With this new formalism, isospin conservation is obtained by setting the value of $M^{2\nu}_{F}$ to zero. The results in ref. \cite{SRFV13} show, that for spherical nuclei this new approach reduces the $0\nu\beta\beta$ Fermi matrix elements by about 30$\sim$40\%, while the more important Gamow-Teller contributions are unchanged. In this work we adopt this new formalism for the $0\nu\beta\beta$-decay of strongly deformed $^{150}$Nd. This is done by separating the renormalization of the particle-particle residual proton-neutron  interaction into the T=1 ($g_{pp}^{T=1}$) and T=0 ($g_{pp}^{T=0}$) parts. The new treatment differs from our work in \cite{FFRS11}, where $g_{pp}^{T=0}=g_{pp}^{T=1}$ and the $g_{pp}$'s are fitted to the experimental two neutrino double beta decay matrix elements $M^{2\nu}_{GT}$. The previous approach \cite{FFRS11}  yields a relative large value of the Fermi part $M^{2\nu}_{F}$, which should disappear according to isospin conservation. In the present work we restore (at least partially) the isospin symmetry for the first time in the QRPA approach for deformed nuclei and calculate the $0\nu\beta\beta$-decay matrix element for the deformed $^{150}$Nd nucleus.

At first we give a brief review of our method. The QRPA states are defined as:
\begin{eqnarray}
|K^\pi,m\rangle=\sum_{pn} \left(X^{m}_{pn} \alpha^\dagger_{p}\alpha^\dagger_{n} - Y^m_{pn}\alpha_{\tilde{n}}\alpha_{\tilde{p}} 
\right) |0\rangle\\
(\Omega_p+\Omega_n=K);\quad \pi_p\pi_n=\pi \nonumber
\end{eqnarray}
Where the $\alpha$'s are the quasi-particle creation and annihilation operators. $K^\pi$ is the angular momentum projection to the symmetry axis of the axially symmetric deformed nucleus. $|0\rangle$ is the QRPA vacuum. But in actual calculations it is always simplified to the  BCS vacuum. X's and Y's are derived from the solutions of the QRPA equations in the deformed system as in ref.  \cite{YRFS08}:
\begin{eqnarray}
\left(
\begin{array}{cc}
A(K) & B(K) \\
-B(K) & -A(K)
\end{array}
\right)
\left(
\begin{array}{c}
X^K \\
Y^K
\end{array}
\right)
=\omega^K_m
\left(
\begin{array}{c}
X^K \\
Y^K
\end{array}
\right)
\end{eqnarray}
Now due to partial isospin restoration  the detailed expressions of $A$ and $B$ for the realistic G-matrix are slightly different from those in \cite{YRFS08}:
\begin{eqnarray}
& &A_{pn,p'n'}(K)=\delta_{pn,p'n'}(E_n+E_p) \nonumber \\
&-&g_{ph}(u_p v_n u_{p'} v_{n'}+v_p u_n v_{p'} u_{n'}) V_{pn'p'n} \nonumber \\
&+& (u_p u_n u_{p'} u_{n'}+v_p v_n v_{p'} v_{n'}) (g_{pp}^{T=0} V_{p\bar{n}p'\bar{n'}}^{T=0} + g_{pp}^{T=1} V_{p\bar{n}p'\bar{n'}}^{T=1})\nonumber \\ 
&&B_{pn,p'n'}(K)=g_{ph}(u_p v_n u_{p'} v_{n'}+v_p u_n v_{p'} u_{n'}) V_{pn'p'n} \nonumber \\
&+& (u_p u_n u_{p'} u_{n'}+v_p v_n v_{p'} v_{n'}) (g_{pp}^{T=0} V_{p\bar{n}p'\bar{n'}}^{T=0} + g_{pp}^{T=1} V_{p\bar{n}p'\bar{n'}}^{T=1})\nonumber \\ 
\end{eqnarray} 
Here the particle-hole interactions $V_{pnp'n'}$ are the same as in \cite{YRFS08}. The particle-particle interactions are now isospin dependent. They are now different for the T=0 and T=1 parts. These two parts are expressed by expansion into the  G-matrix elements in the spherical basis:
\begin{eqnarray}
V_{p\bar{n}p'\bar{n'}}^{T=0(1)}&=&\sum_{J} \sum_{\eta_p\eta_n\eta_{p'}\eta_{n'}} F^{JK}_{p\eta_p\bar{n}\eta_n} F^{JK}_{p'\eta_{p'}\bar{n'}\eta_{n'}} \nonumber \\
&\times&G^{T=0(1)}(\eta_p \eta_n \eta_{p'} \eta_{n'} J)
\label{eq5}
\end{eqnarray}
The decomposition coefficients $F$ are defined in \cite{YRFS08}.

QRPA yields the matrix elements of the $2\nu\beta\beta$- and the $0\nu\beta\beta$-decay as:
\begin{eqnarray}
M^{\beta\beta}_{\mathcal{O}}&=&\sum_{m_i,m_f}^{K^\pi} \langle pnK^\pi| \mathcal{O} |p'n'K^\pi \rangle_{m_f,m_i} \nonumber \\
&\times& \langle 0_f^+ | \widetilde{c_p^\dagger c_n} | K^\pi m_f\rangle 
\langle K^\pi m_f | K^\pi  m_i\rangle \nonumber \\ 
&&\langle K^\pi m_i| c_p^\dagger c_n |0_i^+\rangle 
\label{eq6} 
\end{eqnarray}
For the $2\nu\beta\beta$-decay, the first term in the above formula can be divided in two parts:
\begin{eqnarray}
\langle pnK^\pi| \mathcal{O}^{2\nu}_{GT} |p'n'K^\pi \rangle_{m_f,m_i}&=&\frac{ \langle p |\tau^+{\bf \sigma}| n \rangle \cdot \langle p' |\tau^+{\bf \sigma}|n' \rangle}
{E_{g.s.} +(E_{m_i}+E_{m_f})/2} \nonumber \\
\langle pnK^\pi |\mathcal{O}^{2\nu}_F|p'n'K^\pi \rangle_{m_f,m_i} &=&\frac{ \langle p| \tau^+|n\rangle \langle p'|\tau^+ |n'\rangle }{E_{g.s.}+(E_{m_i}+E_{m_f})/2} \nonumber \\
\label{eq7}
\end{eqnarray}
The detailed expressions of the single particle transition matrix elements $\langle p |\tau^+\sigma| n \rangle $ in the deformed system can be found in \cite{YRFS08,FFRS11}. Similar expressions can be obtained for the F matrices. The denominators are now slightly  different from \cite{FFRS11}. Here $E_{g.s.}=(2M(^{150}$Pm$)-M(^{150}$Nd$)-M(^{150}$Sm$))/2$, and $E_{m}$'s are given in ref.  \cite{FBS13} as  $E_{m}=\omega_m-\omega_{lst}$. The  $\omega$'s are eigenvalues of QRPA and $\omega_{lst}$ are the lowest eigenvalues of QRPA for different $K^\pi$'s. 
For QRPA, the transition matrix elements between the ground states and intermediate states are expressed as:
\begin{eqnarray}
 \langle K^\pi ,m| c_p^\dagger c_n |0^+  \rangle 
 = (u_{p} v_{n}X^{K^\pi m}_{pn} - v_{p} u_{n} Y^{K^\pi m}_{pn}) 
\end{eqnarray}
$u$'s and $v$'s are the occupation amplitudes of the BCS equations. X's and Y's are solutions of the QRPA equations.
The overlap of the initial and final intermediate states in eq. (\ref{eq6}) is more complicated and is given in refs. \cite{SPF03,SPVF99,FFRS11}. The overlap between the two BCS vacua $_f\langle BCS|BCS\rangle_i$ suppresses the matrix elements strongly, if the deformations of the initial and final nuclei are different \cite{SPF03}. We discussed this effect on the matrix element in ref. \cite{FBE11}. In this work we take the above BCS  overlap equal to $0.52$ as given  in Table I of ref. \cite{FFRS11}.

The $0\nu\beta\beta$-decay operators are two body operators with integrations over the loop momentum $q$, see \cite{FFRS11,SPVF99}. The total $0\nu\beta\beta$ matrix element can be expressed as:
\begin{eqnarray}
M'^{0\nu} = \left(\frac{g_A}{1.27}\right)^2\ \left(-M^{0\nu}_F\frac{g^2_V}{g^2_A} + M^{0\nu}_{GT}  \right) 
\label{eq1}
\end{eqnarray}
It contains a Fermi (F) and a Gamow Teller (GT) part and depends on the vector $g_V$ and the axial vector $g_A$ coupling constants. $M^{0\nu}$'s contain summations over different transition operators \cite{SPVF99}. In this work we neglect the tensor contributions. They are usually small compared to the other two parts \cite{SRFV13}.

\begin{table}
\centering
\caption{A brief summary of the parameters used in the current calculations which are changed compared to ref. \cite{FFRS11}. Other parameters are not changed. The last line displays the old parameters used in ref. \cite{FFRS11}, where both $g_{pp}$ values for $T=0$ and $T=1$ are the same and thus isospin is not restored.}
\begin{tabular}{|c|cc|cc|}
\hline
 & \multicolumn{2}{c|}{$N=4-6$} & \multicolumn{2}{c|}{$N=0-7$} \\
&$g_A=1.27$ & $g_A=1.0$ & $g_A=1.27$ & $g_A=1.0$ \\
\hline
$g_{pp}^{T=0}$ &1.08 & 1.03 & 0.77 & 0.74\\
$g_{pp}^{T=1}$ & \multicolumn{2}{c|}{1.34} &\multicolumn{2}{c|}{1.06}  \\
\hline
$g_{pp}$\cite{FFRS11} & 1.05 & 1.00 &  & \\
\hline
\end{tabular}
\label{par}
\end{table}

We now give briefly  the parameters of the model used in this work. The single particle energies and wave functions are obtained by solving the Schr\"odinger equation with a Coulomb corrected Woods-Saxon potential. The same deformation parameters $\beta_2$ , for $^{150}$Nd $\beta_2=0.240$ and $^{150}$Sm $\beta_2=0.153$, as in ref. \cite{FFRS11} are used. The pairing strengths $g_{pair}$'s are obtained by fitting the experimental gaps. The strength of the particle-hole residual interaction $g_{ph}$ is the same as in our earlier description of the  $0\nu\beta\beta$ decay in $^{150}$Nd. The $g_{pp}$ now changes in the T=1 channel $g_{pp}^{T=1}$, see table \ref{par}. They are obtained by fitting the experimental $2\nu\beta\beta$-decay matrix elements. We found a similar behavior as in ref. \cite{SRFV13}: The GT part $M^{2\nu}_{GT}$ depends only on $g_{pp}^{T=0}$ and not on  $g_{pp}^{T=1}$. While the $M_F^{2\nu}$ depends strongly on $g_{pp}^{T=1}$ but not on $g_{pp}^{T=0}$. Thus $g_{pp}^{T=0}$ and $g_{pp}^{T=1}$ can be determined separately by reproducing the experimental GT matrix elements, which for these nuclei are $M^{2\nu}_{GT}=0.07$ \cite{Bar10} and by enforcing $M^{2\nu}_F=0$ as required by isospin symmetry. (The reason that we call this "partial restoration of isospin symmetry" is, that in exact shell model calculations, where the isospin symmetry is conserved, all the successive single particle transitions from initial to intermediate and then to final states are exactly zero. For QRPA only the overall sum of these transitions disappears.)  Ref.\cite{SRFV13} shows, that $g_{pp}^{T=1}$ should be approximately the same as the pairing strength $d_{pp}$ and $d_{nn}$ in the $T=1$ channel. We checked this in the present calculations in the large model space. We find for $^{150}$Nd, $d_{pp}=0.94$ and $d_{nn}=1.03$, and for $^{150}$Sm, $d_{pp}=0.95$ and $d_{nn}=1.04$. This gives an  average pairing strength of $\ {d}=0.98$ and $\bar{d}=1.00$ respectively, about $5\%$ smaller than $g_{pp}^{T=1}$. These results agree with ref.\cite{SRFV13} and imply that the new parameter introduced is consistent to the pairing strength in the T=1 channel.

\begin{figure*}
\includegraphics[scale=0.5]{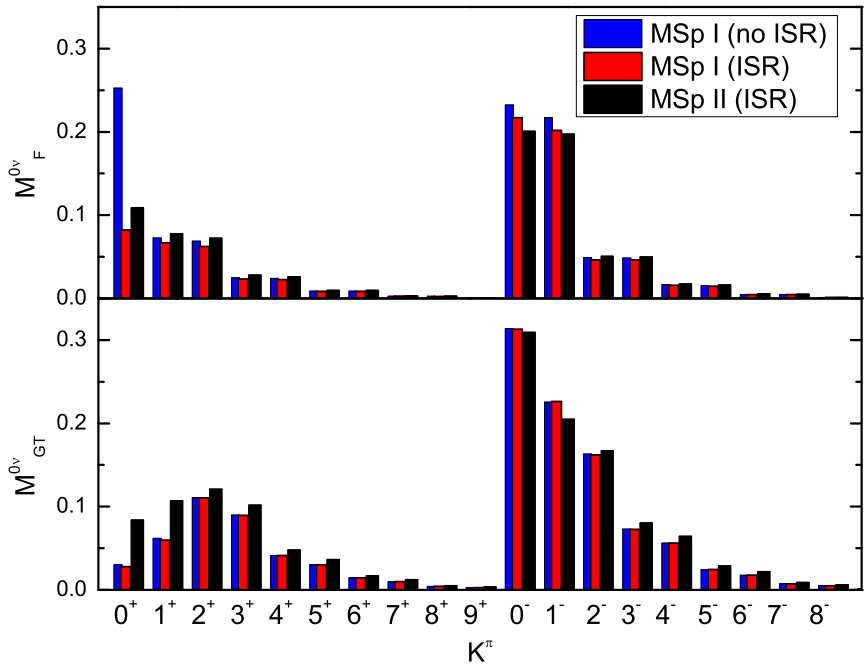}
\caption{(Color online) The decomposition of $0\nu\beta\beta$ matrix elements for different values of  $K^\pi$. Here, "Msp" is an abbreviation for "model space", where "Msp I" refers to the N=4-6 and "Msp II" to the  N=0-7 space. "ISR" abbreviates "Isospin Symmetry Restoration".}
\label{meK}
\end{figure*}

In the present  calculations, we fit two sets of $M^{2\nu}_{GT}$ values, one for the bare $g_{A0}=1.27$ and another for the quenched  $g_A=0.75g_{A0}$ values obtained from experiment \cite{Gue11}. Using calculated $1^+_1$ excitation energies in the energy denominator, barely changes the matrix elements compared to ref. \cite{FFRS11}. The newly fitted $g_{pp}^{T=0}$ values only differ by 1 to 2 \% while $M^{0\nu}_{GT}$ is basically not changed. As a result of improved computational facilities we can now use a much larger model space of up to eight major shells $N=0-7$ compared with a model space $N=4-6$ used in ref.  \cite{FFRS11}. The values of $g_{pp}$ in this larger model space are smaller. This implies that in a smaller model space, the interaction in the particle-particle channel is enhanced to account for the smaller model space. The enlargement of the model space changes the results of $0\nu\beta\beta$-decay as we shall show below.   

We illustrate the results of single intermediate $K^\pi$ contributions in Fig.\ref{meK}. The blue bars are the results in the small basis N = 4 to 6 without isospin symmetry restoration and with the conventional choice of $g_{pp}^{T=0}=g_{pp}^{T=1}$. The symmetry restored results in the small basis are displayed by the red bars. Here as for all results calculated in this paper the CD-Bonn nucleon-nucleon Brueckner G-matrix elements and the Brueckner short range correlations ({\it src}) of ref. \cite{SFM09} are used.

The effect of the isospin restoration leads to changes of $M^{0\nu}_F$ for each $K^\pi$. The largest change is obtained  for  $K^\pi = 0^+$, which corresponds to $J^\pi=0^+$  in the spherical limit in ref.  \cite{SRFV13}. For $K^\pi = 0^+$ the Fermi matrix element  $M^{0\nu}_F$ is reduced to about $1/3$. Changes for other $K^\pi$'s are not that significant. Compared with the conventional parametrization, the new formalism increases $g_{pp}^{T=1}$ by $0.35$ (more than 30\%, see table. \ref{par}).  But this large increase does barely change the values of $M^{0\nu}_{GT}$, since its main contribution is due to T = 0 nucleon pairs.
 Only $M_F^{0\nu}$ for $K^\pi = 0^+$ is sensitive to $g_{pp}^{T=1}$, because the main contribution originates from the interaction in T = 1 nucleon pairs. 

\begin{table}
\centering
\caption{A summary of the results calculated with the Bonn CD potential with different model spaces, with or without Brueckner short range correlations (scr and no scr) of the Bonn CD nucleon-nucleon interaction, two axial vector coupling constants 
$g_A = 1.27$ and $g_A = 0.95$ and with and without partial isospin restoration (ISR, no ISR). The matrix element $M'^{0\nu}$ is defined in eq. (\ref{eq1}). }
\begin{tabular}{|cc|ccc|ccc|}
\hline 
& & \multicolumn{3}{c|}{$g_A=1.27$} & \multicolumn{3}{c|}{$g_A=0.95$}  \\
& & $M_{F}^{0\nu}$ & $M_{GT}^{0\nu}$ & ${M'}^{0\nu}$ & $M_{F}^{0\nu}$ & $M_{GT}^{0\nu}$ & ${M'}^{0\nu}$ \\
\hline
N=4-6 & ISR, no src  & -1.308 & 2.081 & 2.891 & -1.306 & 2.371 & 2.143 \\
 &  no ISR, no src  & -1.565 & 2.091 & 3.061 & -1.614 & 2.381 & 2.340 \\
 & ISR, src        & -1.367 & 2.214 & 3.062 & -1.365 & 2.508 & 2.257 \\
 & no ISR, src  & -1.628 & 2.224 & 3.233 & -1.679 & 2.518 & 2.457 \\
 \hline
N=0-7 & ISR, no src  & -1.390 & 2.309 & 3.171 & -1.369 & 2.629 & 2.328 \\
         & ISR, src & -1.454 & 2.466 & 3.367 & -1.433 & 2.790 & 2.458 \\
\hline
\end{tabular}
\label{res}
\end{table}

\begin{table*}
\centering
\caption{Results for the total matrix element $M'^{0\nu}$ (defined in eq. (\ref{eq1})). $g_A=1.27$ and the self-consistent Brueckner CD-Bonn short range correlations \cite{SFM09} are used. One should be aware different methods use different conventions such as nuclear radii and short range correlations,  {\it etc}, which affect the final results. Here "Non-closure" means, that the intermediate states are calculated explicitly, and "Closure" means, one calculates the transitions from ground states to ground states without taking into account explicitly the intermediate states.}
\begin{tabular}{|cccc|cccc|}
\hline
 \multicolumn{4}{|c|}{Non-closure} & \multicolumn{4}{c|}{Closure} \\
previous \cite{FFRS11} & this work & QRPA-SK\cite{ME13} &LP-QRPA\cite{Ter15}& PHFB\cite{RCCRH10} & IBM-2\cite{BKI15} & NREDF\cite{RM10} & REDF\cite{YSHRM15} \\
\hline
3.34 & 3.37 & 2.71&3.60 & 3.24 & 2.67 & 1.71 &5.60\\
\hline
\end{tabular}
\label{oth}
\end{table*}

In fig. \ref{meK}, we show also how the enlargement of the model space affects the final results. Amazingly previous truncations of the  model space, though numerically insufficient due to its small size, produce however similar results as the ones obtained from a large model space. The main increase of $M^{0\nu}_{GT}$ is due to the two states $K^\pi=0^+$ and $K^\pi= 1^+$ by the larger model space. For other $K^\pi$'s the large model space could either increase or reduce slightly the matrix elements depending on the detailed transitions. This doesn't mean that the contributions from transitions outside of the truncated model space N = 4 to 6 are not important, since we have quite different renormalization strength parameters  $g_{pp}$'s for both T=0 and T=1 parts for the truncated and large model space (see table \ref{par}). 
For the $2\nu\beta\beta$-decay larger values for $g_{pp}$'s are required for the small model space N = 4 to 6, because some of the correlations are missing in the truncated model space and we need to compensate these correlations in the QRPA calculations by increasing the interaction strength. 
The situation for the $0\nu\beta\beta$-decay is similar to the $2\nu$ decay. The larger $g_{pp}$'s in the small space mimic the behavior in a much larger model space and produce values close to the values of the matrix elements in the large space. So the larger force strength $g_{pp}$ compensates for the smaller model space and finally one obtains very similar transition matrix elements for the $0\nu\beta\beta$-decay (see table \ref{par} and ref.  \cite{DLF10}). In general (See fig. 1) the larger model space enhances the transition matrix elements calculated with the short range Brueckner correlations using  the CD-Bonn force ({\it src}, see also table \ref{res}) especially for GT matrix elements. For the Fermi (F) part the situation is different. This  indicates a smaller sensitivity of Fermi matrix elements on the size of the model space.

In table \ref{res}, we have summarized all the results with different model spaces and compared the new formalism with Isospin Restoration (ISR) with results without ISR in the small model space $N = 4 - 6$. The isospin restoration (ISR) reduces the Fermi marix elements $M^{0\nu}_F$ by about $15$ to $20 \%$. The main reduction originates from $K^\pi = 0^+$. 
$M^{0\nu}_{GT}$ is almost not changed (see table \ref{res} box N = 4 - 6). The overall matrix element $M'^{0\nu}$ is reduced by less than $10\%$ including the ISR.  For the dependence on the axial charge $g_A$ the Gamow-Teller matrix element $M^{0\nu}_{GT}$ increases for a decreasing  $g_A$. The result for the total matrix element going from $g_A = 1.27$ to $ 0.95$ is a reduction by about $30\%$. The effect of short range correlations (src)  is the same as in ref. \cite{FFRS11}. The Brueckner short range correlation src  for the Bonn CD force slightly enhance in the small and the large model space the matrix elements by about  $5\%$. The CD-Bonn interaction is known to yield small short range correlation (src). Thus one expects larger changes for other short range correlations.

The results for the absolute values of the Fermi and the Gamow-Teller (GT) matrix elements are enhanced by about $5$ and $11\%$, respectively,  enlarging the model space for both axial charges considered.  
This enhancement, as we have seen from fig. \ref{meK}, stems mostly from the $K^\pi=0^+$ and $K^\pi = 1^+$ ($J^\pi=1^+$) states for the GT part. This is due to the sensitivity on $g_{pp}^{T=0}$ for these  $K^\pi$'s. A detailed analysis of these sensitivity  will be presented in a future work. In refs. \cite{RFSV03} one argues, that in a  spherical system this enhancement will be partially compensated by a decrease of the Tensor part, which is not included in the current calculation. The spherical calculations in ref. \cite{RFSV03} suggests,  that the enhancement due to the enlarged model space could be smaller than shown here. Further investigations are needed for such a conclusion  in a deformed system.

In table \ref{oth}, we compare recent results of $0\nu\beta\beta$-decay for $^{150}Nd$ from different methods. Compared with our previous results, we now have partially restored isospin symmetry, which reduces $M^{0\nu}_F$. Due to the new energy denominator and the enlarged model space, the values are slightly  increased compared to  QRPA calculations based on Skyrme forces \cite{ME13},  which do not include short range correlations. Our results are smaller than that of ref. \cite{Ter15}, which uses pp and nn  QRPA. Except for QRPA most methods use the closure approximation, where one does not need to determine the intermediate states, but treat it by closure. It has been suggested by \cite{SHB14} that this approximation changes the results by at most 10\% for shell model calculations. Some further investigations are needed to see, if this holds for all methods. In table \ref{oth}, we have listed results from closure methods. They deviate strongly from each other. The PHFB gives a final matrix element very close to ours. IBM-2 gives a much smaller value close to Skyrme QRPA. The two results using the energy density functional (EDF) give the largest (5.6, ref. \cite{YSHRM15}) and the smallest values ( 1.71, ref. \cite{RM10}). They are different by a factor 3.  These authors  need to investigate further the strange large difference between the two  "density functional" results.

In conclusion, currently we still have variations of  the $0\nu\beta\beta$-decay matrix elements for different methods by a factor of three or more, which can produce an uncertainty of the order of one magnitude for the half-lives. Further comparisons among methods should be  studied to find  the reason for such large deviations.

\end{document}